\global\def\draftcontrol{0}

   \def\versionno{ relaxation }

\catcode`\@=11

\expandafter\ifx\csname draftcontrol\endcsname\relax\global\def\draftcontrol{0}
\fi

{\count255=\time\divide\count255 by 60
\xdef\hourmin{\number\count255}
\multiply\count255 by-60\advance\count255 by\time
\xdef\hourmin{\hourmin:\ifnum\count255<10 0\fi\the\count255}}
\def\draftdate{\number\month/\number\day/\number\year\ \ \ \hourmin }

\newcommand\makepapertitle{\par
  \begingroup
    \renewcommand\thefootnote{\@fnsymbol\c@footnote}%
    \def\@makefnmark{\rlap{\@textsuperscript{\normalfont\@thefnmark}}}%
    \long\def\@makefntext##1{\parindent 1em\noindent
            \hb@xt@1.8em{%
                \hss\@textsuperscript{\normalfont\@thefnmark}}##1}%
     \newpage
     \global\@topnum\z@   
     \@makepapertitle
     \thispagestyle{empty}\@thanks
  \endgroup
  \setcounter{footnote}{0}%
  \global\let\thanks\relax
  \global\let\makepapertitle\relax
  \global\let\@makepapertitle\relax
  \global\let\@thanks\@empty
  \global\let\@author\@empty
  \global\let\@date\@empty
  \global\let\@title\@empty
  \global\let\title\relax
  \global\let\author\relax
  \global\let\date\relax
  \global\let\and\relax
  \def\version{\let\version\@version\@gobble}
}
\def\@makepapertitle{%
  \newpage
   \ifnum\draftcontrol=1 {}
   \version\versionno
   \vskip 3em%
   \else
   \hfill\hbox to 3cm {\parbox{4cm}{\@pubnum}\hss}%
   \vskip 3em%
   \fi
   \begin{center}%
   \let \footnote \thanks
     {\LARGE {\@title}}%
     \vskip 1.5em%
     {\normalsize
       \lineskip .5em%
       \begin{tabular}[t]{c}%
         \@author
       \end{tabular}\par}%
     \vskip 1.5em%
     {\@bstract}%
     \end{center}%
     \vskip 1.5em
     \@date%
   \par
}

\gdef\@pubnum{}
\def\pubnum#1{%
  \gdef\@pubnum{#1}}

\gdef\@bstract{}
\def\Abstract#1{%
  \gdef\@bstract{%
   \parbox{\textwidth-0pc}{%
   \centerline{\bf Abstract}\penalty1000%
\kern.2cm%
\noindent
\renewcommand\baselinestretch{1.0}%
{#1}}}
}

\def\ps@paper{\let\@mkboth\@gobbletwo%
     \ifnum\draftcontrol=1
    \def\@oddfoot{\hbox to \textwidth{\tiny \versionno \hfil\tiny\draftdate}%
    \hskip -\textwidth \hbox to \textwidth{\hfil\rm\thepage\hfil}}%
     \else\def\@oddfoot{\hbox to \textwidth{\hfil\rm\thepage\hfil}}
     \fi
     \let\@evenfoot\@oddfoot
}

\def\body{\clearpage
          \pagestyle{paper}
    }

\def\@version#1{\ifnum\draftcontrol=1
\typeout{}\typeout{#1}\typeout{}
\vskip3mm\centerline{\hbox{\fbox{\normalsize{\tt DRAFT -- #1 -- }
                   {\draftdate}}}}\vskip3mm
\fi}
\let\version\@version
\long\def\eqlabel#1{\ifnum\draftcontrol=1
                    \tag@false  
                    \tag*{(\theequation) \hbox to -0.2cm{\hspace{0cm}\small{#1}\hss}}
                    \refstepcounter{equation}
                    \edef\@currentlabel{\theequation}
                    \ltx@label{#1}          
                    \else
                    \label{#1}
                    \fi
                    }
\let\st@bibitem\@bibitem
\let\st@lbibitem\@lbibitem
\ifnum\draftcontrol=1
  \def\@bibitem#1{%
    \st@bibitem{#1}\a@@label{#1}\ignorespaces}
  \def\@lbibitem[#1]#2{%
    \st@lbibitem[#1]{#2}\a@@label{#2}\ignorespaces}
  \def\a@@label#1{%
    \gdef\a@lab{\smash{\normalfont\small#1}}
    \ifvmode
      \if@inlabel
        \global\setbox\@labels\hbox{%
          \llap{\a@lab\let\a@lab\relax
                \kern\@totalleftmargin\kern\marginparsep}%
          \box\@labels}%
      \fi
    \fi}
\fi

\documentclass[12pt,letterpaper]{article}

\usepackage{amsmath,amssymb,array,calc,epsfig,rotating,bm}
\usepackage[sort]{cite}
\usepackage{graphicx}
\usepackage{psfrag,verbatim}

\ifnum\draftcontrol=1
\fi

\renewcommand\baselinestretch{1.25}
\renewcommand\section{\@startsection {section}{1}{\z@}%
                                   {-3.5ex \@plus -1ex \@minus -.2ex}%
                                   {2.3ex \@plus.2ex}%
                                   {\normalfont\large\bfseries}}
\renewcommand\subsection{\@startsection{subsection}{2}{\z@}%
                                   {-3.25ex\@plus -1ex \@minus -.2ex}%
                                   {1.5ex \@plus .2ex}%
                                   {\normalfont\normalsize\bfseries}}
\renewcommand\subsubsection{\@startsection{subsubsection}{3}{\z@}%
                                   {-3.25ex\@plus -1ex \@minus -.2ex}%
                                   {1.5ex \@plus .2ex}%
                                   {\normalfont\normalsize\it}}
\renewcommand\paragraph{\@startsection{paragraph}{4}{\z@}%
                                   {-3.25ex\@plus -1ex \@minus -.2ex}%
                                   {1.5ex \@plus .2ex}%
                                   {\normalfont\normalsize\bf}}

\numberwithin{equation}{section}

\def\revise#1       {\raisebox{-0em}{\rule{3pt}{1em}}%
                     \marginpar{\raisebox{.5em}{\vrule width3pt\
                     \vrule width0pt height 0pt depth0.5em
                     \hbox to 0cm{\hspace{0cm}{%
                     \parbox[t]{4em}{\raggedright\footnotesize{#1}}}\hss}}}}

\def\cale         {{\cal E}}

\def\caln         {{\cal N}}
\def\calo         {{\cal O}}
\def\calp         {{\cal P}}

\def\del          {\partial}

\def\sqr#1#2{{\vcenter{\vbox{\hrule height.#2pt
 \hbox{\vrule width.#2pt height#1pt \kern#1pt
 \vrule width.#2pt}\hrule height.#2pt}}}}

\newcommand{\ft}[2]{{\textstyle{\frac{#1}{#2}}}}

\def\b{\beta}

\newcommand{\beq}{\begin{equation}}
\newcommand{\eeq}{\end{equation}}
\newcommand{\beqa}{\begin{eqnarray}}
\newcommand{\eeqa}{\end{eqnarray}}
\newcommand{\beqar}{\begin{eqnarray*}}
\newcommand{\eeqar}{\end{eqnarray*}}

\renewcommand{\eqref}[1]{(\ref{#1})}

\newcommand{\ie}{{\it i.e.,}\ }

\def\w{\omega}

\def\t{\tau}

\catcode`\@=12

\begin{document}


\title{\bf Relaxation time of non-conformal plasma}
\pubnum{UWO-TH-09/13}

\date{August 2, 2009}

\author{
Alex Buchel\\[0.4cm]
\it Perimeter Institute for Theoretical Physics\\
\it Waterloo, Ontario N2L 2Y5, Canada\\[.5em]
 \it Department of Applied Mathematics\\
 \it University of Western Ontario\\
\it London, Ontario N6A 5B7, Canada
 }

\Abstract{We study effective relaxation time of viscous hydrodynamics of strongly coupled non-conformal gauge theory 
plasma using gauge theory/string theory correspondence. We compute leading corrections to the conformal plasma relaxation time from the
relevant deformations due to dim-2 and dim-3 operators.  We discuss in details the relaxation time $\t_{eff}$ of $\caln=2^*$ plasma.  
For a certain choice of masses this theory undergoes a phase transition with divergent specific heat $c_V \sim|1-T_c/T|^{-1/2}$.
Although the bulk viscosity remains finite all the way to the critical temperature, we find  that $\tau_{eff}$ diverges 
near the critical point as $\tau_{eff}\sim |1-T_c/T|^{-1/2}$. 
}

\makepapertitle

\body

\version\versionno
\tableofcontents

\section{Introduction}
Understanding the properties of strongly coupled quark-gluon plasma (sQGP) produced at RHIC \cite{shuryak} is an active research area.
Of particular interest is the development of viscous hydrodynamic codes describing the evolution of the plasma ball at 
early stages of heavy ion collisions \cite{c1,c2,c3}. These hydrodynamic simulations require input of the phenomenological parameters, 
such as the shear $\eta$ and the bulk $\zeta$ viscosities, as well as higher order transport coefficients necessary to maintain causality 
of hydrodynamic description \cite{m,is,higher1,higher2} (and typically, numerical stability of the codes). Currently, reliable 
computation of transport 
coefficients of sQGP is not available. As a result, one resorts to analysis of exactly soluble models of strongly coupled 
gauge theory plasma. Transport properties of a large class of strongly coupled gauge theory plasmas can be studied in a 
framework of gauge theory/string theory correspondence \cite{juan,review1}. Here, one discovers universal relations for the 
(first order) transport coefficients  \cite{u1,u2,u3,u4}
\begin{equation}
\frac{\eta}{s}=\frac{1}{4\pi}\,,\qquad \frac{\zeta}{\eta}\ge 2 \left(\frac 13- c_s^2\right)\,, 
\eqlabel{order1}
\end{equation} 
where $s$ is the entropy density and $c_s$ is the speed of sound in the plasma. 
The hope is that it is generic relations of the type \eqref{order1} that might provide an input into realistic simulations of sQGP.

In this letter we study a specific second-order transport coefficient ---  the effective relaxation time $\t_{eff}$ --- in 
strongly coupled non-conformal gauge theory plasma. The motivation for our analysis is the recent work of Song and 
Heinz \cite{sh} where it was suggested that the relaxation time near the phase transition can be much larger due to 
``critical slowing down''. Thus the sQGP produced in heavy ion collisions at RHIC might have strong 
sensitivity to the initial 
conditions for the bulk viscous pressure.  

We begin in the next section with introducing the notion of the effective relaxation time. The simplest way to obtain 
non-conformal viscous 4-dimensional hydrodynamics is to compactify a conformal hydrodynamics in $(4+k)$-dimensions on a flat 
$k$-dimensional 
torus \cite{u4}. We discuss effective relaxation time of such compactified plasma. In section \ref{rgfixed} we compute corrections 
to the effective relaxation time in the vicinity of the renormalization group (RG) fixed point induced by relevant operators 
(the mass terms). In section \ref{n2} we present results of the detailed analysis of the effective relaxation time 
in $\caln=2^*$ plasma. We focus on a specific mass deformation of $\caln=4$ supersymmetric Yang-Mills (SYM) plasma which 
leads to a phase transition in the infrared. We summarize our results in section \ref{disc}.

\section{$\t_{eff}$ from conformal hydrodynamics}\label{cft}
The most general causal viscous relativistic hydrodynamics has many second order transport coefficients: 5 -  
in the case of conformal hydrodynamics \cite{higher1}, and 13 - in the case of non-conformal hydrodynamics \cite{rom1}. 
It is impractical to simulate hydrodynamic evolution in the full multiparameter phenomenological space of these transport
coefficients. Thus, one typically limits discussion to a single 'effective' second order transport coefficient. 
Since different groups use different approximations of the general viscous hydrodynamics, it is necessary to related the analysis 
through some physical observable. We propose to use a dispersion relation of the linearized sound waves in plasma
to define a common effective relaxation time $\t_{eff}$:
\begin{equation}
\w=\pm c_s k-i \Gamma k^2\pm \frac{\Gamma}{c_s}\biggl(c_s^2 \t_{eff}-\frac{\Gamma}{2}\biggr)k^3+\calo(k^4)\,,
\eqlabel{deft}
\end{equation}  
where $c_s$ is the speed of the sound waves (obtained from the equation of state), and $\Gamma$ is the sound wave attenuation
(determined by the shear and the bulk viscosities)
\begin{equation}
c_s^2=\frac{\del\calp}{\del\cale}\,,\qquad \Gamma=\left(\frac 23 \frac{\eta}{\cale+\calp}+\frac 12 \frac{\zeta}{\cale+\calp}\right)\,.
\eqlabel{csg}
\end{equation} 
The relaxation time $\t_{eff}$ is the relaxation time of M\"uller-Israel-Stewart hydrodynamics \cite{m,is}; 
it is the relaxation time of relativistic conformal 
hydrodynamics as defined in \cite{higher1}. In general non-conformal hydrodynamics \cite{rom1}, $\t_{eff}$ is a combination of 
particular second-order transport coefficients\footnote{See \cite{rom1} for definition of $\{\tau_\pi,\t_\Pi\}$.}:
\begin{equation}
\t_{eff}=\frac{\t_\pi+\ft 34 \ft{\zeta}{\eta}\ \t_\Pi}{1+\ft 34 \ft{\zeta}{\eta}}\,.
\eqlabel{teffr}
\end{equation}

In $\caln=4$ SYM plasma at infinite 't Hooft coupling and in the planar limit, the relaxation time is \cite{higher1,higher2}\footnote{Finite 
coupling/non-planar corrections to the relaxation time of conformal plasmas are discussed in \cite{bp,bhm}. }
\begin{equation}
\tau_{eff}T\equiv  \t_\pi^\star T= \frac{2-\ln 2}{2\pi} \,.
\eqlabel{n4r}
\end{equation}

Starting from the conformal hydrodynamics in $d=(4+k)$-dimensions we can obtain non-conformal hydrodynamics by  compactification on
a flat $k$-torus  \cite{u4}. Since compactification can not modify the long-wavelength dispersion relation, the hydrodynamic 
dispersion relation for the sound waves \eqref{deft} must be the same. In other words, if $\{c_s^{(d)},{\Gamma}^{(d)},\tau_{\pi}^{(d)}\}$
are the hydrodynamic coefficients of the $d$-dimensional conformal plasma, then
\begin{equation}
c_s={c_s}^{(d)}=\frac{1}{\sqrt{d-1}}\,,\qquad \Gamma={\Gamma}^{(d)}=\frac{1}{4\pi T}\ \frac{d-2}{d-1}\,,\qquad \tau_{eff}=\t_{\pi}^{(d)}\,,
\eqlabel{tor1}
\end{equation}
where we used an explicit expression for the attenuation in $d$-dimensional conformal hydrodynamics \cite{higher1} 
\begin{equation}
\Gamma^{(d)}=\frac{d-2}{d-1}\ \frac{\eta^{(d)}}{\cale^{(d)}+\calp^{(d)}}\,,
\eqlabel{tor2}
\end{equation}
the CFT equation of state $\cale^{(d)}= (d-1)\calp^{(d)}$, and the universality of the shear viscosity to the entropy density ratio in strongly coupled plasmas 
\cite{u1,u2,u3}. Given \eqref{csg}, it is easy to see that the bulk viscosity bound proposed in \cite{u4} is saturated in 
such models:
\begin{equation}
\frac{\zeta}{\eta}=2\left(\frac 13-c_s^2\right)\,.
\eqlabel{saturation}
\end{equation}
Notice that the effective relaxation time does not change upon compactification. This fact, the bulk viscosity ratio \eqref{saturation}, 
as well as similar relations for all the other second-order transport coefficients of non-conformal hydrodynamics of \cite{rom1} 
can be obtained by a direct reduction of $d$-dimensional conformal hydrodynamics of \cite{higher1} on a flat $k$-torus. 
For example, it can be shown that the ``bulk'' $\tau_\Pi$ and the ``shear'' $\t_\pi$ relaxation times are the same 
(see also \cite{rom1})
\begin{equation}
\t_\Pi=\t_\pi=\t_\pi^{(d)}\,.
\eqlabel{same}
\end{equation}  
The relaxation time of the $d=5$ holographic CFT plasma was computed in \cite{hy}, and of the $d=6$ holographic CFT plasma in \cite{no}:
\begin{equation}
\begin{split}
\tau^{(5)}_\pi T=&\frac{5}{8\pi}\left(2-\frac{\pi}{5}\sqrt{1-\frac{2}{\sqrt{5}}}+\frac {1}{\sqrt{5}}\ \coth^{-1}\sqrt{5}-\frac 12\ \ln 5\right)\,,\\
\t^{(6)}_\pi T=&\frac{3}{4\pi}\left(2-\frac{\pi}{6\sqrt{3}}-\frac 12\ \ln 3\right)\,.
\end{split}
\eqlabel{taud}
\end{equation}  
Notice that in both cases $d=\{5,6\}$, 
\begin{equation}
\frac{\t_\pi^{(d)}}{\t_{eff}}>1\,.
\eqlabel{trat}
\end{equation}
Eq.~\eqref{trat} is our first indication that the effective relaxation time in non-conformal strongly coupled plasmas  
is {\it larger} than that of the four-dimensional conformal plasma. In the following sections we encounter more examples of this phenomenon.

\section{$\t_{eff}$ in the vicinity of the renormalization group fixed point}\label{rgfixed}
An RG fixed point is a conformal theory. The gauge/gravity correspondence  \cite{juan} not only allows us 
to study a non-trivial (strongly interactive) fixed points (such as $\caln=4$ SYM at large 't Hooft coupling), but  
also provides tools of exploring the deformations of such fixed points by relevant (in the infrared) operators (see \cite{ps} 
for a pedagogical discussion). In this section we discuss the relaxation time of the strongly coupled holographic plasma 
in the vicinity of the fixed point 
perturbed by dimension-2 and dimension-3 operators. Controllable fixed points that are realized in the holographic framework are those 
of the supersymmetric gauge theories. In these cases we can think of the deformation operators as mass terms for the 
gauge theory bosons and fermions. Thermodynamic and hydrodynamic properties of mass deformed conformal gauge theories 
(in the holographic setting) were extensively studied in \cite{n21,n22,n23,n24,u4,n25}. In particular, it is straightforward to 
extend analysis of \cite{n23} and compute the $\calo(k^3)$ contribution to the sound wave dispersion relation \eqref{deft}.
We can then extract the effective 
relaxation time $\t_{eff}$. It is convenient to parametrize the result in terms of the deviation parameter $\delta$,
\begin{equation}
\delta\equiv \frac 13-c_s^2\,,
\eqlabel{defd}
\end{equation} 
away from the fixed point:
\begin{equation}
\t_{eff}=\t_\pi^\star \biggl(1+\beta_{[d]}\ \delta+\calo(\delta^2)\biggr)\,,
\eqlabel{taunon}
\end{equation}
where $\t_\pi^\star$ is the universal relaxation time of the holographic conformal hydrodynamics at (infinitely) strong coupling \eqref{n4r}, 
and $\beta_{[p]}$ is the correction induced by the operator with ${\rm dim}[\calo]=p$. Explicitly we find,
\begin{equation}
\b_{[p]}=
\begin{cases} $2.2837(0)$\,, &\ \text{$p=3$\,,}
\\
$6.3016(8)$\,, &\ \text{$p=2$}\,.
\end{cases}
\eqlabel{cases}
\end{equation}
Again, in both cases the effective relaxation time is larger than that of the conformal plasma.

\section{$\t_{eff}$ in $\caln=2^*$ plasma}\label{n2}
We would like now to study the effective relaxation time in a holographic model with strongly broken scale invariance. 
Our choice is the $\caln=2^*$ gauge theory at strong coupling. Recall that $\caln=2^*$ gauge theory is obtained as 
a supersymmetric mass deformation of $\caln=4$ SYM theory.  The low-energy effective description of this theory is
 exactly soluble \cite{dw}; furthermore, for a specific point on its Coulomb branch one can construct 
an explicit holographic dual \cite{pw} and verify complete agreement between the field-theoretic and the dual gravitational 
descriptions \cite{bpp,ejp}. At a finite temperature, one has additional freedom of assigning different masses $m_b\ne m_f$ to the 
bosonic and the fermionic components of the $\caln=2$ hypermultiplet \cite{n21}.  The thermodynamics of this theory was studied extensively 
in \cite{n24}, and the first order transport coefficients in \cite{n25}. Here, we extend the analysis of \cite{n25} for 
the dispersion relation of the sound waves in $\caln=2^*$ plasma to order $\calo(k^3)$. Using \eqref{deft}, we can extract the 
corresponding $\t_{eff}$.   
\begin{figure}[t]
\begin{center}
\psfrag{d}{{$\left(\frac 13-c_s^2\right)$}}
\psfrag{t}{{$3c_s^2\times \frac{\t_{eff}}{\t_\pi^\star}$}}
  \includegraphics[width=5in]{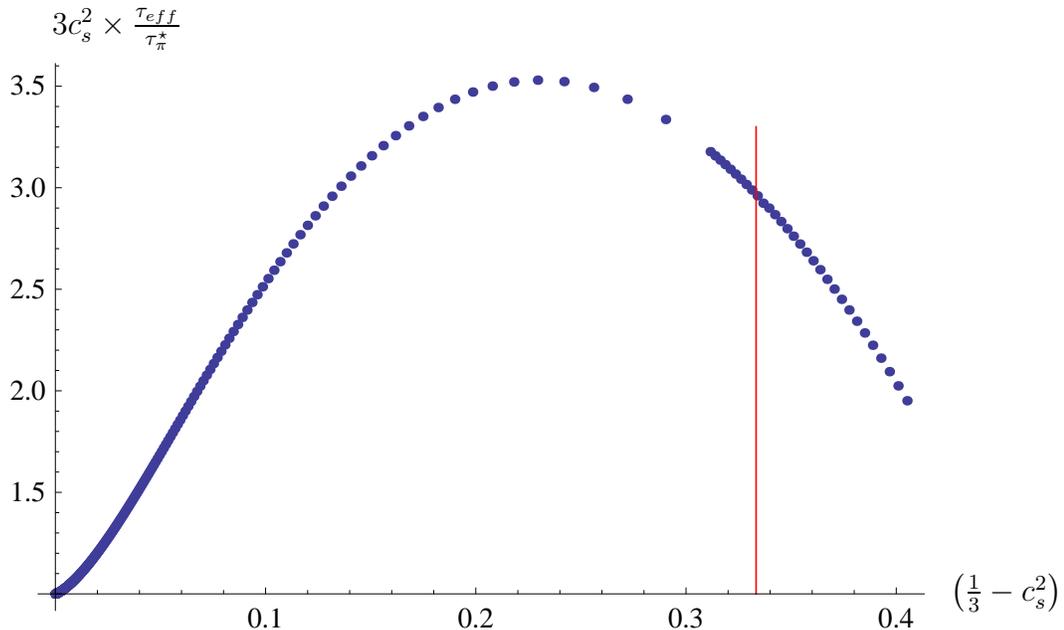}
\end{center}
  \caption{
(color online) Effective relaxation time $\t_{eff}$ of $\caln=2^*$ strongly coupled plasma. The vertical red line indicates 
a phase transition with vanishing speed of sound. }\label{figure1}
\end{figure}

In what follows, we focus on a specific $\caln=2^*$ mass deformation of $\caln=4$ SYM, namely the one with $m_f=0\,,\ m_b\ne 0$.
This gauge theory undergoes an interesting phase transition for $T<T_c\approx m_b/2.29(9)$, where it  
becomes unstable with respect to energy density fluctuations \cite{n24}.
Specifically, precisely at $T=T_c$ the speed of sound waves squared $c_s^2$ vanishes. 
A perturbative instability of this type is a defining feature of a second order phase transition. Since $c_s^2\propto (T-T_c)^{1/2}$,
the specific heat $c_V$ diverges as $|1-T_c/T|^{-1/2}$, suggesting that  
such a critical point is in the universality class of the  mean-field tricritical point\footnote{An identical 
critical point was recently found in the strongly coupled cascading gauge theory plasma \cite{bcas}.}. 
Much like real QCD \cite{kl}, the $\caln=2^*$ plasma has a plateau in the reduced energy density $\frac{\cale}{T^4}$ which extends 
almost to the phase transition - at $T\sim \frac 12 m_b$ the $\caln=2^*$ equation of state deviates from the conformal 
equation of state by less than $3\%$ \cite{n24}. $\caln=2^*$ plasma has a rapidly growing bulk viscosity in the vicinity of the 
phase transition: at the transition point \cite{u4,n25},
\begin{equation}
\frac{\zeta}{\eta}\bigg|_{T=T_c}=6.65(3)\,, \qquad {\rm or}\qquad  \frac{\zeta}{s}\bigg|_{T=T_c}=0.52(9)\,,
\end{equation}  
which is close to the peak value $\frac {\zeta}{s}\sim 0.7$ extracted from lattice QCD \cite{meyer}.

Techniques for determining  the sound wave dispersion relation in strongly coupled 
 $\caln=2^*$ plasma has been developed in 
\cite{n23,n25}. We refer the reader for computational details to the original work. There is one subtlety though: 
in analyzing the differential equations describing the graviton wave function at order $\calo(k^2)$ in the momentum 
(and correspondingly its dispersion relation at order $\calo(k^3)$), we observe that these equations become singular 
in the limit $c_s^2\to 0$. The latter singularity is removed if instead of directly extracting $\t_{eff}$ 
we compute $c_s^2 \t_{eff} $ (along with appropriate rescaling of the graviton wave function). 
Upon such rescaling\footnote{A similar procedure has been employed in \cite{u4,n25}.}, 
all the quantities remain finite in the vicinity of small 
$c_s^2$. The results of the computations are presented in Figure~\ref{figure1}.  
Given that the speed of sound in $\caln=2^*$ plasma is always less than the conformal value \cite{n24}, 
\ie $\frac 13$, 
we see that  for all temperatures down to $T_c$
\begin{equation}
\t_{eff}> \t_{\pi}^\star\,.
\eqlabel{res1}
\end{equation}
Moreover, since the quantity 
$c_s^2 \t_{eff}$ remains finite at $T=T_c$ while the speed of sound vanishes as $c_s^2\propto (T-T_c)^{1/2}$,
we conclude that $\t_{eff}$ actually diverges as $T\to T_c$ 
\begin{equation}
\t_{eff} T_c\ \propto |1-T_c/T|^{-1/2}\,,\qquad T\to T_c+0\,.
\eqlabel{res2}
\end{equation}

\section{Conclusion}\label{disc}
In this letter we studied effective relaxation time in non-conformal strongly coupled plasmas using holographic techniques.
We discussed theories obtained by compactification of higher-dimensional conformal plasmas on flat tori, theories obtained 
by deforming a conformal theory with a relevant operator. Finally, using $\caln=2^*$ strongly coupled plasma as 
an example, we explored the effective relaxation time in the regime with strongly broken scale invariance. 
In all cases we observed that effective relaxation time is longer (and often much longer) than the relaxation 
time of the $\caln=4$ SYM plasma in the planar limit and at infinite 't Hooft coupling \cite{higher1,higher2}. 
In fact, in the $\caln=2^*$ model we find that the relaxation time diverges near the phase transition with vanishing speed of 
sound\footnote{The same general features of the effective relaxation time can be extracted from 
the phenomenological models of holographic gauge/gravity correspondence studied in \cite{ts}.
I would like to thank Todd Springer for sharing the results of his analysis.
}.  
This example suggests that the critical slowing down in sQGP is a distinct possibility \cite{sh}. Further 
studies of more realistic models of QCD are necessary to settle this issue. From the holographic perspective, 
the relaxation time in the cascading gauge theory \cite{ks} near the deconfinement phase transition should 
provide a useful estimate.

In the derivation of the second order viscous hydrodynamics from Boltzmann equations one finds that  
the effective relaxation time is \cite{bolt}
\begin{equation}
\t_{eff}^{Boltzmann} T=\frac{3\eta T}{2\calp }=6\ \frac{\eta}{s}\ \gtrsim\ \frac {3}{2\pi}\,,
\eqlabel{boltz}
\end{equation}   
which is  much larger than $\t_\pi^\star T$ since the ratio of the shear viscosity to the entropy density 
at weak coupling substantially exceed the KSS viscosity bound \cite{u2}.
This fact, supplemented with the analysis of the relaxation time at strong coupling presented in this letter,
might lead to the speculation that $\t_\pi^\star$ is the universal lower bound on the relaxation time.  
Unfortunately, much like it is the case with the KSS bound \cite{vio1,vio2,vio3}, this is not so:
even though the finite 't Hooft coupling corrections tend to increase the relaxation time \cite{bp},
the non-planar corrections would generically decrease the relaxation time \cite{bhm}. 
Moreover, in a class of conformal gauge theory plasmas discussed in \cite{bm}, the microscopic causality
of the theory imposes the bound 
\begin{equation}
\t_{eff}\ge 0.56(1)\ \t_\pi^\star\,.
\eqlabel{taubound}
\end{equation}

\section*{Acknowledgments}
I would like to thank  Rob Myers, Paul Romatschke and Todd Springer for interesting discussions and correspondence.
Research at Perimeter Institute is supported by the Government of
Canada through Industry Canada and by the Province of Ontario
through the Ministry of Research \& Innovation. I gratefully
acknowledges further support by an NSERC Discovery grant and support
through the Early Researcher Award program by the Province of
Ontario.

\end{document}